\documentclass[aps,pre,twocolumn,showpacs,showkeys,superscriptaddress]{revtex4}
\usepackage{epsfig,color}
\usepackage{times}

\begin{document}

\title{Evolutionary dynamics of group interactions on structured populations: A review}

\author{Matja\v{z} Perc}
\email{matjaz.perc@uni-mb.si}
\affiliation{Faculty of Natural Sciences and Mathematics, University of Maribor, Koro\v{s}ka cesta 160, SI-2000 Maribor, Slovenia}

\author{Jes{\'u}s G{\'o}mez-Garde{\~n}es}
\affiliation{Department of Condensed Matter Physics, University of Zaragoza, E-50009 Zaragoza, Spain}
\affiliation{Institute for Biocomputation and Physics of Complex Systems (BIFI), University of Zaragoza, E-50018 Zaragoza, Spain}

\author{Attila Szolnoki}
\affiliation{Institute of Technical Physics and Materials Science, Research Centre for Natural Sciences, Hungarian Academy of Sciences, P.O. Box 49, H-1525 Budapest, Hungary}

\author{Luis M. Flor{\'i}a}
\affiliation{Department of Condensed Matter Physics, University of Zaragoza, E-50009 Zaragoza, Spain}
\affiliation{Institute for Biocomputation and Physics of Complex Systems (BIFI), University of Zaragoza, E-50018 Zaragoza, Spain}

\author{Yamir Moreno}
\affiliation{Institute for Biocomputation and Physics of Complex Systems (BIFI), University of Zaragoza, E-50018 Zaragoza, Spain}
\affiliation{Department of Theoretical Physics, University of Zaragoza, E-50009 Zaragoza, Spain}
\affiliation{Complex Networks and Systems Lagrange Lab, Institute for Scientific Interchange, Viale S. Severo 65, I-10133 Torino, Italy}

\begin{abstract}
Interactions among living organisms, from bacteria colonies to human societies, are inherently more complex than interactions among particles and nonliving matter. Group interactions are a particularly important and widespread class, representative of which is the public goods game. In addition, methods of statistical physics have proven valuable for studying pattern formation, equilibrium selection, and self-organisation in evolutionary games. Here we review recent advances in the study of evolutionary dynamics of group interactions on structured populations, including lattices, complex networks and coevolutionary models. We also compare these results with those obtained on well-mixed populations. The review particularly highlights that the study of the dynamics of group interactions, like several other important equilibrium and non-equilibrium dynamical processes in biological, economical and social sciences, benefits from the synergy between statistical physics, network science and evolutionary game theory.
\end{abstract}

\pacs{02.50.-r, 87.23.-n, 89.65.-s, 89.75.-k}

\keywords{evolution, cooperation, public goods, phase transitions, pattern formation, cyclical interactions, self-organization, lattices, complex networks, coevolution, statistical physics}

\maketitle

\section*{1. INTRODUCTION}
We present a review of recent advances on the evolutionary dynamics of spatial games that are governed by group interactions. The focus is on the public goods game, or more generally $N$-player games, which are representative for this type of interaction. Although relevant aspects of $2$-player games are surveyed as well, we refer to \cite{nowak_06} for a more thorough exposition. Another important aspect of this review is its focus on structured populations. In the continuation of this introductory section we will also summarise basic results concerning the public goods game on well-mixed populations, but we refer the reader to \cite{sigmund_10, archetti_jtb12} for details.

The methodological perspective that permeates throughout the review is that of statistical physics. The advances reviewed therefore ought to be of interest to physicists that are involved in the interdisciplinary research of complex systems, but hopefully also to experts on game theory, sociology, computer science, ecology, as well as evolution and modeling of socio-technical systems in general. Group interactions are indeed inseparably linked with our increasingly interconnected world, and thus lie at the interface of many different fields of research. We note that there are many studies which are not covered in this review. However, we have tried to make it as comprehensive as possible to facilitate further research.

We describe our motivation, notation, and other elementary concepts in subsection 1.1, followed by ``in a nutshell'' survey of results on well-mixed populations in subsection 1.2, and an overview of the organisation of the review in subsection 1.3.

\subsection*{1.1. Motivation and basic concepts}
Given that fundamental interactions of matter are of pairwise nature, the consideration of $N$-particle interactions in traditional physical systems is relatively rare. In computational approaches aimed at modeling social, economical and biological systems, however, where the constituents are neither point mass particles nor magnetic moments, $N$-player interactions are almost as fundamental as $2$-player interactions. Most importantly, group interactions in general cannot be reduced to the corresponding sum of pairwise interactions.

\begin{figure}
\begin{center}
\includegraphics[width =5.0cm]{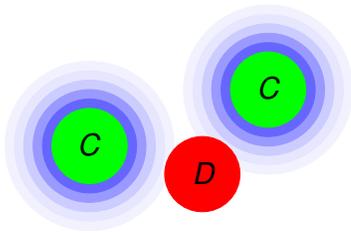}
\caption{Schematic configuration of $N=3$ microbes, where a fraction $\rho_C = 2/3$ are cooperators (producers) and $\rho_D = 1/3$ are defectors (free-riders). For the most popular choice of benefit and cost functions, $\beta(\rho_C) = r \rho_C$ ($r>1$) and $\alpha(\rho_C)=1$ respectively, individual payoffs are $P_C(2/3) = 2r/3 -1$ and $P_D(2/3) = 2r/3$. An explicit computation of $P_C(\rho_C)$ ($1/3 \leq \rho_C \leq 1$) and $P_D(\rho_C)$ ($0\leq \rho_C \leq 2/3$) reveals that they cannot be generated by means of pairwise interactions, thus illustrating the inherent irreducibility of group interactions.}
\label{fig_scheme}
\end{center}
\end{figure}

A simple model inspired by experiments \cite{chuang_sci09, gore_n09, chuang_msb10, frey_pa10, schuster_bs11, archetti_jtb11, cremer_srep12} can be invoked for both motivating the usage of the public goods game as well as for introducing basic notation. Consider a colony of $N$ microbial agents where a fraction of them (producers or cooperators) pour amounts of a fast diffusive chemical into the environment. The latter has the status of a public good as it is beneficial also for those that do not produce it (free-riders or defectors). For $N=3$ such a setup is depicted schematically in Fig.~\ref{fig_scheme}. The metabolic expenses stemming from the production cost of the public good are given by the cost function $\alpha(\rho_C)$, while the individual benefit for each of the $N$ microbes is $\beta(\rho_C)$, where $0 \leq \rho_C \leq 1$ is the fraction of producers. Each non-producing ($D$-phenotype) microbe thus receives the payoff $P_D = \beta(\rho_C)$, while each microbe that does produce ($C$-phenotype) bears the additional cost, so that its net benefit is $P_C =  \beta(\rho_C) - \alpha(\rho_C)$.

For $N=2$ and the simple choice of $\alpha(\rho_C)=c/(2\rho_C)$ and $\beta(\rho_C)=b\theta(\rho_C)$ [where $\theta(x)$ is the step function], we recover two well-known games that are governed by pairwise interactions. Namely the prisoner's dilemma for $2b>c > b>0$ and the snowdrift game for $b > c>0$, as summarized in Table~\ref{table}. When $N \geq 3$, however, the problem becomes that of group interactions. We see that, under the sensible assumption of additivity of individually obtained payoffs, the defined payoff structure cannot be reproduced by means of pairwise interactions (see caption of Fig.~\ref{fig_scheme} for details). This example also suggests that, provided the benefit and cost functions could be inferred from experiments, the experimenter could potentially determine whether a colony is governed by pair or group interactions. Indeed, it was recently noted \cite{archetti_jtb12} that the oversimplifying restriction of pairwise social interactions has dominated the interpretation of many biological data that would likely be much better interpreted in terms of group interactions.

The payoffs $P_D = \beta(\rho_C)$ and $P_C =  \beta(\rho_C) - \alpha(\rho_C)$ have a general public goods game structure in that cooperators bare an additional cost besides the benefits that are common to both strategies. The analysis of decision making by a ``rational microbe'' thus falls within the realm of classical game theory. In this framework, for a constant individual production cost $\alpha$ and an arbitrary concave benefit function $\beta(\rho_C)$, Motro \cite{motro_jtb91} showed that even values from within the $0<\rho_C<1$ interval are stable Nash equilibria. Under certain conditions to be met by the benefit and cost functions ($\beta$ and $\alpha$), there is thus no ``tragedy of the commons'' \cite{hardin_g_s68}. This may be welcomed news for the liberal (``invisible hand'') supporters of public goods systems: The tragedy of the commons is rationally avoidable even without the ``cognitive'' or ``normative'' capacities required for the existence of additional strategies. Nevertheless, the ``tragedy of the commons'' does occur in the majority of other cases (e.g., linear benefit function $\beta$), where no production of the public good is the only rational individual choice.

\subsection*{1.2. Evolutionary game dynamics}
Turning back to microbial populations, under the assumption that the reproductive power of each microbe is proportional to the net metabolic benefit enjoyed, one arrives at a formal description for the time evolution of the fraction of producers $\rho_C$. This is the realm of evolutionary game dynamics, which implements Darwinian natural selection of phenotypes in populations under frequency-dependent fitness conditions \cite{nowak_06, hofbauer_88, gintis_00, sigmund_10}, as well as in related though non-genetic social and economic systems. In the latter, ``social learning'' assumptions may lead to a very similar evolutionary dynamics provided simple assumptions concerning the cognitive capabilities of agents are accepted.

\begin{table}
\begin{center}
\begin{tabular}{l | c c} \hline \hline
& $C$ & $D$ \\ \cline{2-3}
$C$ & $\;\;R= b- (c/2)\;\;$ & $\;\;S = b-c\;\;$ \\
$D$ & $T=b$ & $P = 0$ \\ \hline \hline
\end{tabular}
\end{center}
\caption{Payoff matrix of $2$-player games if $\alpha=c/(2\rho_C)$ and $\beta=\theta(\rho_C)b$. For $2b >c > b >0$ we have the prisoner's dilemma, and for $b > c >0$ the snowdrift game.}
\label{table}
\end{table}

A calculation that invokes a standard well-mixed population setting (see below and \cite{archetti_ev11, archetti_jtb12, cressman_jtb12}) leads to the differential equation for the expected value $x = \langle \rho_C \rangle$ of the fraction of producers
\begin{equation}
\dot{x} = x (1-x) [W_C(x) - W_D(x)]\;,
\label{replicator}
\end{equation}
where $W_{C,D}(x) $ is the average payoff per either a cooperative or defective individual. This is the replicator equation, which is nonlinear already for linear payoffs. Depending further on the additional properties of $\alpha(x)$ and $\beta(x)$, its analysis may thus be all but straightforward.

Theorems relate the asymptotic states of the replicator equation with Nash's stability criteria, and Motro's \cite{motro_jtb91} results on the public goods game in turn translate into the characterization of the evolutionary stable states for our microbial population. In particular, for constant $\alpha$ and concave benefit functions $\beta(x)$, a well-mixed colony of mixed phenotypic composition is evolutionary stable. Notably, in addition to the replicator dynamics, best-response and related learning dynamics can also be formulated for the evolution of $x$, and indeed they can be of much relevance in specific contexts of agent based modeling.

At this point, it is informative to spell out the operational assumptions that traditionally underlie the well-mixed approximation \cite{archetti_jtb12}. In particular, it is assumed that the $N-1$ individuals that interact with the focal player are randomly sampled from an infinite population of cooperators and defectors, so that the probability of interacting with $j$ cooperators is given by $f_j(x) = C^{N-1}_{j} x^j (1-x)^{N-1-j}$, where $x$ ($1-x$) is the average fraction of cooperators (defectors). Other formally more sophisticated settings can also be of interest. We refer to \cite{cressman_jtb12} for one that allows to consider a continuum of strategies parameterized by the amount of public good produced per individual, and to \cite{pena_evol12} for a ``grand canonical'' treatment where the group size $N$ is considered a random variable.

Note that the payoffs of the focal individual are collected from group configurations that are statistically uncorrelated. Moreover, in order to implement the assumption that the individual reproductive power is proportional to the net benefits, while operationally keeping a constant population size $N$, one can (among other options, like using the stochastic birth/death Moran process) use a replicator-like rule in which, in the next time step, the focal player imitates the current strategy of a randomly chosen agent from the group, with a probability depending on the payoff difference. It is worth emphasising that the basic underlying assumption here is homogeneity, so that individuals do not differentiate or assort, as both (i) the payoff's are collected from, and (ii) the competitive reproduction is against configurations sampled from an unbiased (uncorrelated) strategic distribution $f_j(x)$.

If departing from the assumptions of well-mixed populations, however, several issues open up. To begin with:
\begin{itemize}
\item Which criterion determines how group configurations are sampled to provide instantaneous payoff to focal players? Is the group size $N$ also a random variable in that sampling?
\item What kind of population sampling is used to implement replicating competition among strategies? In other words, who imitates whom? Should members of all groups be potential imitators (or should potentially be imitated)? Or should just a fraction of them (for example those in a smaller spatial neighborhood of the focal player) qualify as such?
\end{itemize}

There are several possible answers to both groups of questions, and they depend significantly on the particular problem one wishes to address. For example, for a quantitative modeling of a yeast colony of invertase-producers and non-producing cells, the answers should be based on considerations involving characteristic time scales of many biochemical processes and the spatial microbial arrangements that are typical amongst the measured samples of the microbial colony, to name but a few potentially important issues. On the other hand, in systems where best-response or other non-imitative evolutionary rules are considered, only the first group of questions would likely be of relevance.

Recent research concerning public goods games on structured populations is in general very indirectly, if at all, related to a particular experimental setup. Instead, it is of an exploratory nature over different potentially relevant theoretical issues that can be either formulated or understood as lattice or network effects. A quite common ground motivation is the search for analogs of network reciprocity \cite{nowak_n92b}. Is the resilience of cooperative clusters against invading defectors on networks and lattices enough to effectively work against the mean-field tendencies \cite{floria_pre09}? More generally, what are the effects of structure in a population when confluent with known sources of public goods sustainability, such as punishment or reward? Are these synergistic confluences? Indeed, the interest of reviewed research goes far beyond its relevance to a specific experiment. Evolutionary game dynamics is of fundamental interest to the making of interdisciplinary complex systems science, encompassing biological, economical as well as social sciences, and from this wider perspective, the universal features of dynamical processes of group interactions are still rather unexplored.

\subsection*{1.3. Organisation of the review}
The remainder of this review is organised as follows. In Sec.~2 we survey the implementation of the public goods game on lattices. We focus on recent studies investigating the effects of lattice structure on the emergence of cooperation. In addition, we review both the effects of heterogeneity in the dynamical ingredients of the public goods game as well as the effects of strategic complexity on the evolution of cooperation. In Sec.~3 we focus on structures that are more representative for human societies. In this framework we will revisit the formulation of the public goods game on complex networks and show how social diversity promotes cooperation. In addition, we will survey how public goods games on networks can be formulated by means of a bipartite representation. The latter includes both social as well as group structure, thus opening the path towards a more accurate study of group interactions in large social systems. We conclude Sec.~3 by reviewing different networked structures in which the public goods game has been implemented, most notably modular and multiplex networks, as well as populations of mobile agents. In Sec.~4 we review advances on structured populations where the connections coevolve with the evolutionary dynamics, and where thus the topology of interactions changes depending on the payoffs and strategies in the population. We round off the review by discussing the main perspectives, challenges and open questions in Sec.~5, and by summarising the conclusions in Sec.~6.

\section*{2. LATTICES}

\begin{figure}
\begin{center}
\includegraphics[width = 4.0cm]{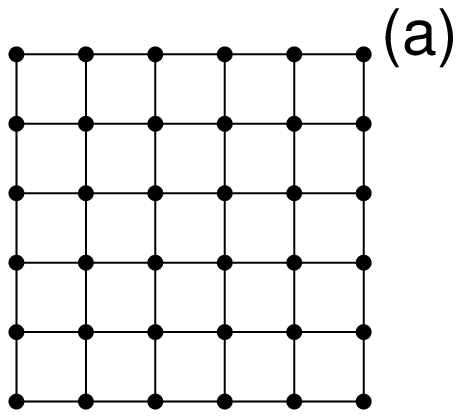} \includegraphics[width = 4.0cm]{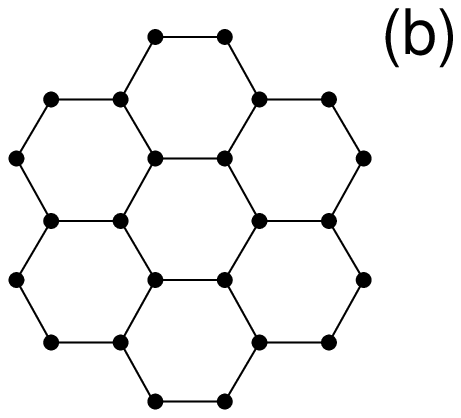}\\
\includegraphics[width = 4.0cm]{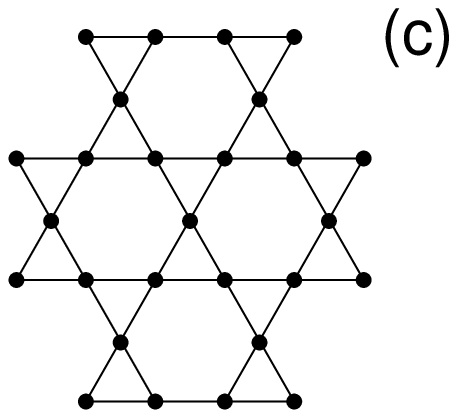} \includegraphics[width = 4.0cm]{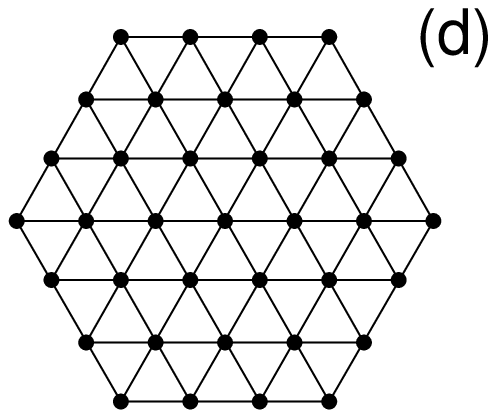}
\caption{Schematic presentation of different types of lattices. On the square lattice (a) each player has four immediate neighbors, thus forming groups of size $G=5$, while on the honeycomb lattice (b) it has three, thus $G=4$. In both cases the clustering coefficient ${\cal C}$ is zero. Yet the membership of unconnected players in the same groups introduces effective links between them, which may evoke behavior that is characteristic for lattices with closed triplets \cite{szolnoki_pre09c}. On the other hand, the kagom{\'e} (c) and the triangular lattice (d) both feature percolating overlapping triangles, which makes them less susceptible to effects introduced by group interactions. The kagom{\'e} lattice has $G=5$ and ${\cal C}=1/3$, while the triangular lattice has ${\cal C}=2/5$ and $G=7$.}
\label{lattices}
\end{center}
\end{figure}

Beyond patch-structured populations where under certain updating rules the spatial structure has no effect on the evolution of altruism \cite{taylor_evo00, irwin_tpb00, irwin_tpb01}, lattices represent very simple topologies, which enjoy remarkable popularity in game theoretical models \cite{nowak_n92b, szabo_pr07, roca_plr09}. Despite their dissimilarity to actual social networks \cite{wasserman_94}, they provide a very useful entry point for exploring the consequences of structure on the evolution of cooperation. Moreover, there are also realistic systems, especially in biology and ecology, where the competition between the species can be represented adequately by means of a lattice \cite{kessler_pre93, iwasa_ee98}. In general, lattices can be regarded as an even field for all competing strategies where the possibility of network reciprocity is given \cite{nowak_n92b}. Furthermore, as there are many different types of lattices (see Fig.~\ref{lattices} for details), we can focus on very specific properties of group interactions and test what is their role in the evolutionary process.

The basic setup for a public goods game with cooperators and defectors as the two competing strategies on a lattice can be described as follows. Initially, $N \propto L^2$ players are arranged into overlapping groups of size $G$ such that everyone is surrounded by its $k=G-1$ neighbors and belongs to $g=G$ different groups, where $L$ is the linear system size and $k$ the degree (or coordination number) of the lattice. Each player on site $i$ is designated either as a cooperator (C) $s_i= 1$ or defector (D) $s_i = 0$ with equal probability. Cooperators contribute a fixed amount $a$, normally considered being equal to $1$ without loss of generality, to the common pool while defectors contribute nothing. Finally, the sum of all contributions in each group is multiplied by the synergy factor $r$ and the resulting public goods are distributed equally amongst all the group members. The payoff of player $i$ in every group $g$ is
\begin{equation}
P_{i}^g=r \frac{\sum_{j\in g} s_j a}{G} -  s_i a =r\frac{ N_C^g a}{G}- s_i a\;,
\end{equation}
where $N_{C}^g$ is the number of cooperators in group $g$. The net payoff $i$ thereby acquires is the sum of the payoffs received in all the groups it participates in: $P_{i} = \sum_{g} P_{i}^g$.

The microscopic dynamics involves the following elementary steps. First, a randomly selected player $i$ plays the public goods game as a member of all the $g=1, \ldots, G$ groups.  Next, player $i$ chooses one of its neighbors at random, and the chosen player $j$ also acquires its payoff $P_{j}$ in the same way. Finally, player $i$ enforces its strategy $s_i$ onto player $j$ with some probability determined by their payoff difference. One of the possible choices for this update probability is the Fermi function,
\begin{equation}
\Pi(s_i \to s_j)=\frac{1}{1+\exp[(P_{j}-P_{i})/GK]}\;,
\label{eq:Fermi}
\end{equation}
where $K$ quantifies the uncertainty by strategy adoptions and $G$ normalises it with respect to the number and size of the groups. These elementary steps are repeated consecutively, whereby each full Monte Carlo step (MCS) gives a chance for every player to enforce its strategy onto one of the neighbors once on average. Alternatively, synchronous updating can also be applied so that all the players play and update their strategies simultaneously, but the latter can lead to spurious results, especially in the deterministic $K \to 0$ limit \cite{huberman_pnas93}. Likewise, as anticipated above, there are several ways of how to determine when a strategy transfer ought to occur, yet for lattices the Fermi function can be considered standard as it can easily recover both the deterministic as well as the stochastic limit. The average fraction of cooperators $\rho_{C}$ and defectors $\rho_{D}$ in the population must be determined in the stationary state. Depending on the actual conditions, such as the proximity to extinction points and the typical size of the emerging spatial patterns, the linear system size has to be between $L=200$ and $1600$ in order to avoid accidental extinction, and the relaxation time has to exceed anywhere between $10^4$ and $10^6$ MCS to ensure the stationary state is reached. Exceptions to this basic requirements are not uncommon, especially when considering more than two competing strategies, as we will emphasise at the end of this section.

\subsection*{2.1. Group versus pairwise interactions}

\begin{figure}
\begin{center} \includegraphics[width = 7.5cm]{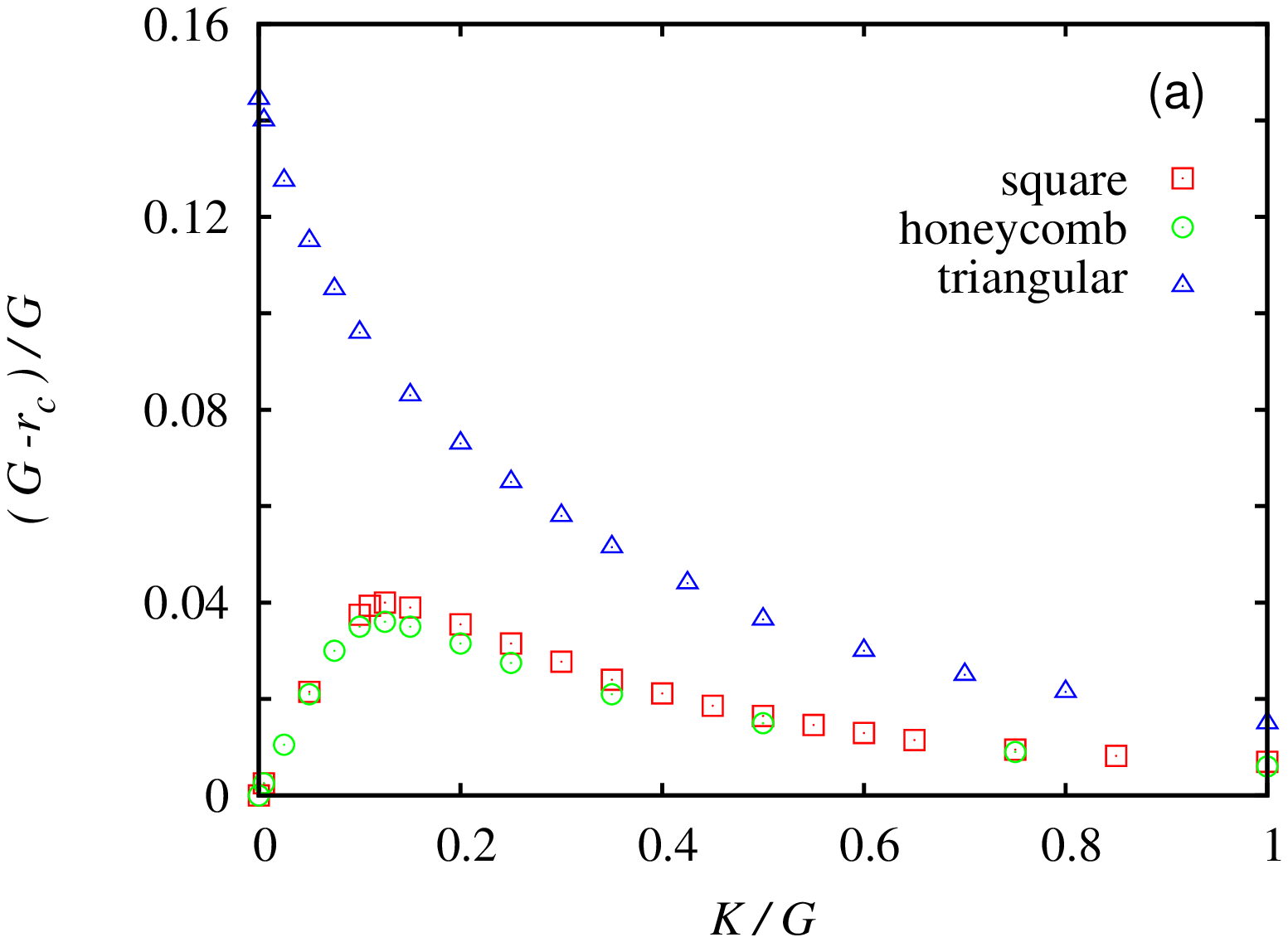}
\includegraphics[width = 7.5cm]{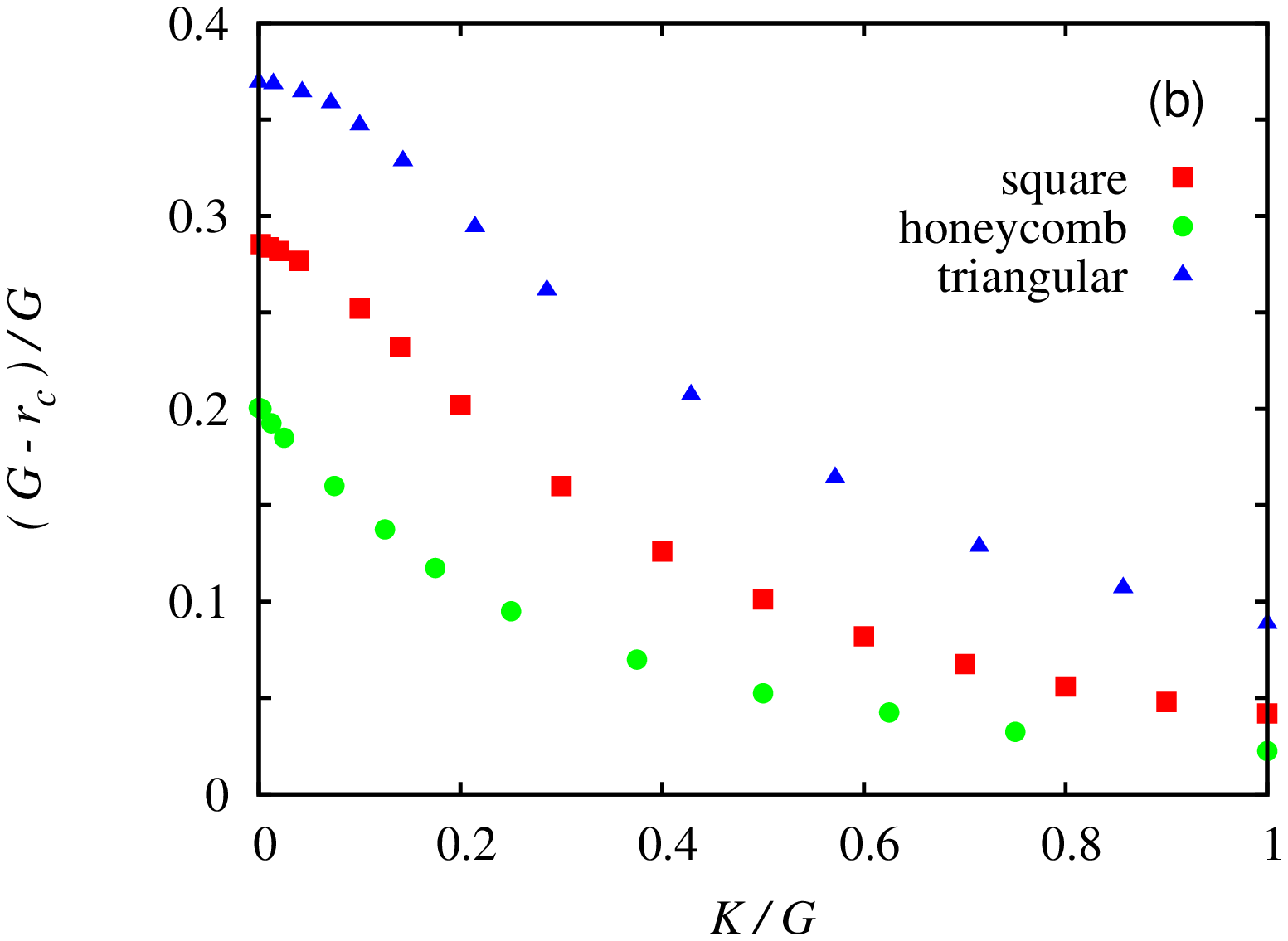}
\caption{Borders between the mixed $C+D$ and the pure $D$ phase in dependence on the normalized uncertainty by strategy adoptions $K/G$, as obtained on different lattices for pairwise (a) and group (b) interactions. Vertical axis depicts the defection temptation rate, i.e., the higher its value the smaller the value of $r$ that still allows the survival of at least some cooperators. By pairwise interactions ($G=2$), the absence of overlapping triangles is crucial (square and honeycomb lattice), as then there exists and intermediate value of $K$ at which the evolution of cooperation is optimally promoted. If triangles do percolate (triangular lattice), the $K \to 0$ limit is optimal. This behavior is characteristic for all social dilemmas that are based on pairwise games, most famous examples being the prisoner's dilemma and the snowdrift game (see Fig.~3 and 5 in \cite{szabo_pre05}). Conversely, when group interactions are considered (see Fig.~\ref{lattices} for $G$ values) the topological differences between the lattices become void. Accordingly, the deterministic $K \to 0$ limit is optimal regardless of the topology of the host lattice \cite{szolnoki_pre09c}.}
\label{effective}
\end{center}
\end{figure}

For games governed by pairwise interactions, such as the prisoner's dilemma game, the dependence of the critical temptation to defect $b_{c}$ on $K$ is determined by the presence of overlapping triangles. Notably, here $b_{c}$ is the temptation to defect $b$ above which cooperators are unable to survive (see also Table~\ref{table}). If an interaction network lacks overlapping triangles, and accordingly has the clustering coefficient ${\cal C} = 0$, as is the case for the square and the honeycomb lattice, then there exists an intermediate $K$ at which $b_{c}$ is maximal. On the other hand, if overlapping triangles percolate, as is the case for the triangular and the kagom{\'e} lattice (see Fig.~\ref{lattices}), then the deterministic limit $K \to 0$ is optimal for the evolution of cooperation \citep{szabo_pre05, vukov_pre06}. The spatial public goods game behaves differently, highlighting that group interactions are more than just the sum of the corresponding number of pairwise interactions. As demonstrated in \cite{szolnoki_pre09c}, group interactions introduce effective links between players that are not directly connected by means of the interaction network. Topological differences between lattices therefore become void, and the deterministic limit $K \to 0$ becomes optimal for the evolution of cooperation regardless of the type of the interaction network. Results for pairwise and group interactions are summarised in Fig.~\ref{effective}. This implies that by group interactions the uncertainty by strategy adoptions plays at most a side role, as it does not influence the outcome of the evolutionary process in a qualitative way.

\begin{figure}
\begin{center} \includegraphics[width = 8.5cm]{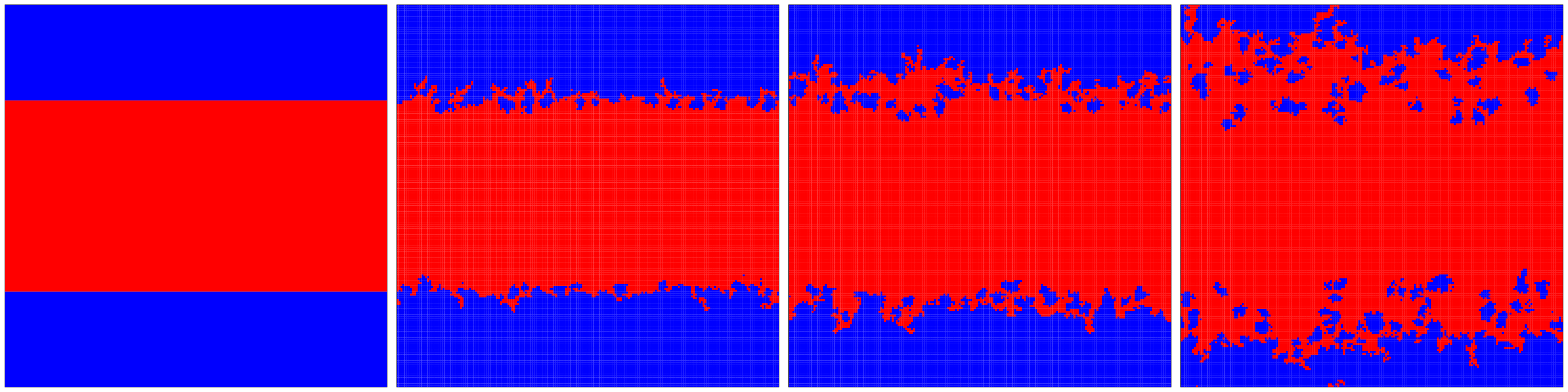} \\
\includegraphics[width = 8.5cm]{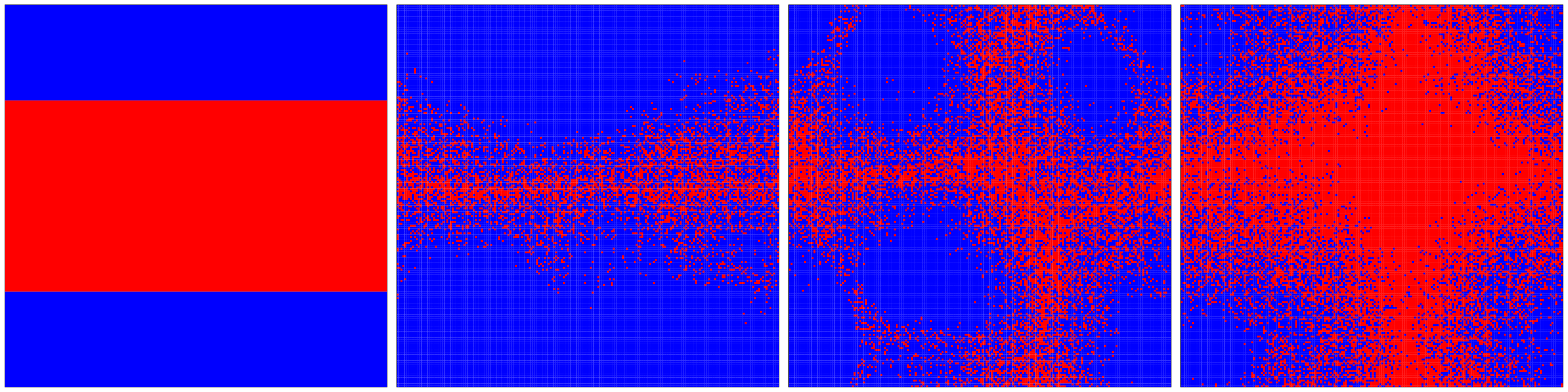}
\caption{Characteristic snapshot of evolutionary process for small (up, $G=5$) and large (down, $G=301$) groups. Cooperators are depicted blue while defectors are depicted red. For small groups, the evolution of strategies proceeds with the characteristic propagation of the fronts of the more successful strategy (in this case $D$) until eventually the maladaptive strategy $C$ goes extinct.
For large groups, however, the cooperative clusters are strong and can outperform the defectors, even if $r$ is very small. Still, as the density of defectors decreases, their payoff suddenly becomes very competitive, and thus they can invade the seemingly invincible cooperative clusters.
Such an alternating time evolution is completely atypical and was previously associated with cooperators only.}
\label{group_size}
\end{center}
\end{figure}

The fact that membership in the same groups effectively connects players that are not linked by means of direct pairwise links naturally brings forth the group size as a key system parameter. In \cite{szolnoki_pre11c} it was shown that increasing the group size does not necessarily lead to mean-field behavior, as is traditionally observed for games governed by pairwise interactions \cite{szabo_pre09}, but rather that public cooperation may be additionally promoted by means of enhanced spatial reciprocity that sets in for very large groups where individuals have the opportunity to collect payoffs separately from their direct opponents. However, very large groups also offer very large benefits to invading defectors, especially if they are rare, and it is this back door that limits the success of large groups to sustain cooperation and puts a lid on the pure number-in-the-group effect \cite{isaac_qje88}. Figure~\ref{group_size} features two characteristic snapshots and further details to that effect. It is also worth emphasising that the joint membership in large groups will indirectly link vast numbers of players, thus rendering local as well global structural properties of interaction networks practically irrelevant for the final outcome of the public goods game.

\subsection*{2.2. Heterogeneities in the dynamics}

Group interactions on structured populations are thus different from the corresponding sum of pairwise interactions. Consequently not just the group size, but also the distribution of payoff within the groups becomes important. As shown in \cite{shi_dm_epl10, perc_njp11}, heterogeneous payoff distributions do promote the evolution of cooperation in the public goods game, yet unlike as games governed by pairwise interactions \cite{perc_pre08}, uniform distributions outperform the more heterogeneous exponential and power law distributions. The setup may also be reversed in that not the payoffs but rather the contributions to the groups are heterogeneous. In \cite{gao_j_pa10, vukov_jtb11} it was shown that correlating the contributions with the level of cooperation in each group markedly promotes prosocial behavior, although the mechanism may fail to deliver the same results on complex interaction networks where the size of groups is not uniform. Conceptually similar studies with likewise similar conclusions are also \cite{cao_xb_pa10, peng_d_epjb10, lei_c_pa10}, although they rely on differences in the degree of each player to determine payoff allocation. The latter will be reviewed in Sec.~3 where the focus is on public goods games that are staged on complex networks. Another possibility to introduce heterogeneity to the spatial public goods game is by means of different teaching activities of players, as was done in \cite{guan_pre07}. In this case, however, the results are similar to those reported previously for games governed by pairwise interactions \cite{szolnoki_epl07}, in that there exists an optimal intermediate density of highly active players at which cooperation thrives best.

\begin{figure}
\begin{center} \includegraphics[width = 7cm]{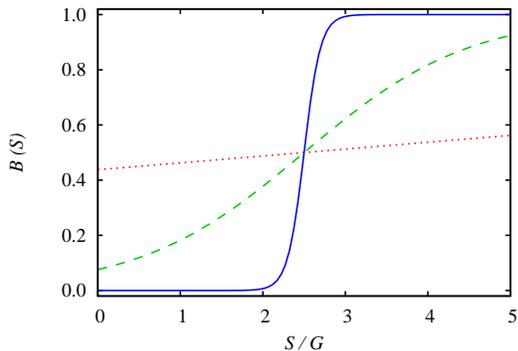}
\caption{Different realizations of the public benefit function $B(S)=\frac{1}{1+\exp[-\beta(S_i-T)]}$, where $T$ represents the threshold value and $\beta$ the steepness of the function \cite{boza_bmceb10}. For $\beta=0$ the benefit function is a constant equalling $0.5$, in which case the produced public goods are insensitive to the efforts of group members. Conversely, for $\beta=+\infty$ the benefit function becomes step-like so that group members can enjoy the benefits of collaborative efforts via $r$ only if the total amount of contributions in the group $S$ exceeds a threshold. Otherwise, they obtain nothing. The depicted curves were obtained for $T=2.5$ and $\beta=0.1$ (dotted red), $1$ (dashed green) and $10$ (solid blue).}
\label{public_benefit}
\end{center}
\end{figure}

\begin{figure}
\begin{center} \includegraphics[width = 8.3cm]{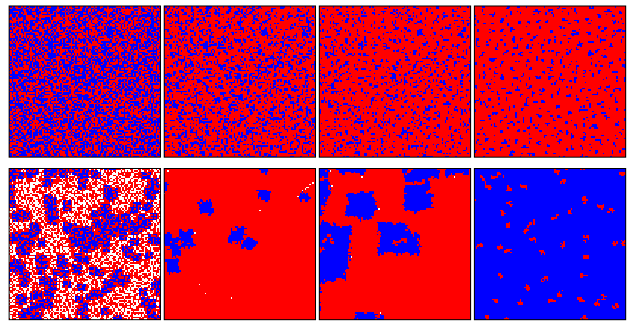}
\caption{Time evolution of strategies on a square lattice having $G=25$, for the critical mass $M = 2$ (up) and $M = 17$ (down) at $r/G=0.6$ \cite{szolnoki_pre10}. Defectors are marked red, while cooperators are depicted blue if their initial contributions are exalted or white if they go to waste. Accordingly, cooperators can be designated as being either ``active'' or ``inactive''. When
$M$ is low all cooperators are active, yet they don't have a strong incentive to aggregate because an increase in their density will not elevate their fitness. Hence, only a moderate fraction of cooperators coexists with the prevailing defectors in the stationary state. If the critical mass is neither small nor large, the status of cooperators varies depending on their location on the lattice: there are places where their local density exceeds the threshold and they can prevail against defectors. There are also places where the cooperators are inactive because their density is locally insufficient and loose against defectors. The surviving domains of active cooperators start spreading, ultimately rising to near dominance.}
\label{critical_mass}
\end{center}
\end{figure}

Aside from heterogeneous distributions of payoffs and initial investments, group interactions are also amenable to different public benefit functions, as demonstrated in Fig.~\ref{public_benefit}. While traditionally it is assumed that the production of public goods is linearly dependent on the number of cooperators within each group, it is also possible to use more complex benefit functions. The idea has been explored already in well-mixed populations \cite{wang_j_pre09, boza_bmceb10, archetti_ev11, deng_k_pone11b}, and in structured populations the possibilities are more. One is to introduce a critical mass of cooperators that have to be present in a group in order for the collective benefits of group membership to be harvested \cite{szolnoki_pre10}. If the critical mass is not reached, the initial contributions can either go to waste, or they can also be depreciated by applying a smaller multiplication factor in that particular group \cite{shi_dm_jsm11, shi_dm_pa12}. Although such models inevitably introduce heterogeneity in the distribution of payoffs \cite{wang_j_pre10}, they can also lead to interesting insights that go beyond ad hoc introduced heterogeneity. In \cite{szolnoki_pre10}, for example, it was shown that a moderate fraction of cooperators can prevail even at very low multiplication factors if the critical mass $M$ is minimal. For larger multiplication factors, however, the level of cooperation was found to be the highest at an intermediate value of $M$. Figure~\ref{critical_mass} features two characteristic scenarios. Notably, the usage of nonlinear benefit functions is unique to group interactions, and in general it works in favor of public cooperation \cite{shi_dm_jsm11, shi_dm_pa12}.

\subsection*{2.3. Strategic complexity}
Besides heterogeneity in payoffs and nonlinearity in public benefit functions, introducing strategic complexity is another way of bringing the public goods game closer to reality. As noted above, the willingness to cooperate may depend on the behavior of others in the group. Correlating the contributions with either the level of cooperation in each group \cite{gao_j_pa10, vukov_jtb11} or the degree of players \cite{cao_xb_pa10, peng_d_epjb10,lei_c_pa10} can thus be seen not just as heterogeneous contributing, but also as conditional cooperation \cite{fischbacher_el01}. An explicit form of this was studied in \cite{szolnoki_pre12}, where conditional cooperator of the type $C_j$ only cooperate provided there are at least $j$ other cooperators in the group. It was shown that such strategies are the undisputed victors of the evolutionary process, even at very low synergy factors. Snapshots of the spatial grid reveal the spontaneous emergence of convex isolated ``bubbles'' of defectors that are contained by inactive conditional cooperators. While the latter will predominantly cooperate with the bulk of active conditional cooperators, they will certainly defect in the opposite direction, where there are defectors. Consequently, defectors cannot exploit conditional cooperators, which leads to a gradual but unavoidable shrinkage of the defector quarantines. Notably, conditional strategies introduced in this way have no impact on the mixed state in unstructured populations and are thus of interest only on structured populations.

\begin{figure}
\begin{center} \includegraphics[width = 8.5cm]{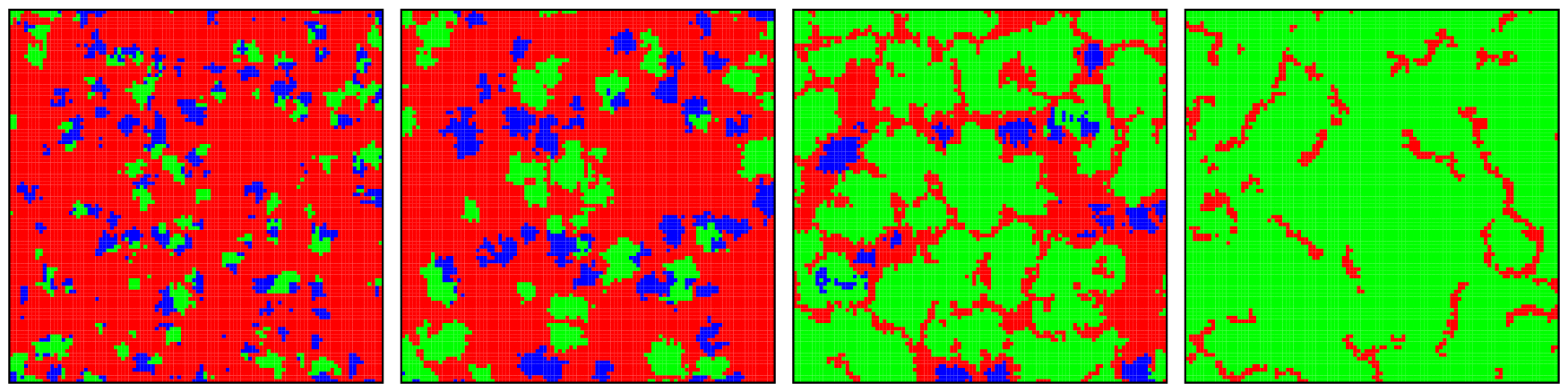} \\
\includegraphics[width = 8.5cm]{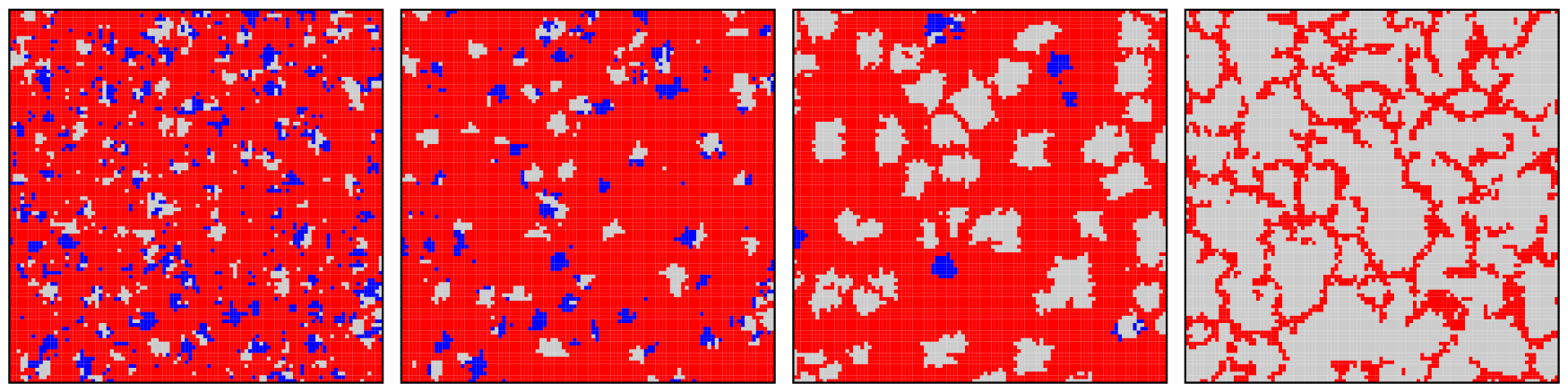}
\caption{Indirect territorial battle between pure cooperators (blue) and peer-punishers (green) (upper row), and between pure cooperators (blue) and rewarding cooperators (light gray) (lower row). In the upper row pure cooperators and peer-punishers form isolated clusters that compete against defectors (red) for space on the square lattice. Since peer-punishers are more successful in competing against defectors than pure cooperators (also frequently referred to as second-order free-riders \cite{fehr_n04}), eventually the latter die out to a leave a mixed two-strategy phase (peer-punishers and defectors) as a stationary state (see \cite{helbing_ploscb10} for further details). In the lower row defectors are quick to claim supremacy on the lattice, yet pure and rewarding cooperators both form isolated compact clusters to try and prevent this. While rewarding cooperators can outperform defectors, pure cooperators can not. Accordingly, the latter die out, leaving qualitatively the same outcome as depicted in the upper row (see \cite{szolnoki_epl10} for further details).}
\label{indirect}
\end{center}
\end{figure}

Apart from conditional strategies, the impact of loners, sometimes referred to as volunteers, has also been studied in the realm of the spatial public goods game \cite{szabo_prl02}. While in well-mixed populations volunteering leads to cyclic dominance between the three competing strategies \cite{hauert_s02, semmann_n03}, on lattices the complexity of the emerging spatial patterns enables the observation of phase transitions between one-, two-, and three-strategy states \cite{szabo_prl02}, which fall either in the directed percolation universality class \cite{hinrichsen_ap00} or show interesting analogies to Ising-type models \cite{liggett_85}.

The complexity of solutions in spatial public goods games with three or more competing strategies is indeed fascinating, which can be corroborated further by results reported recently for peer-punishment, \cite{helbing_ploscb10, helbing_njp10, helbing_pre10c}, pool-punishment \cite{szolnoki_pre11, perc_srep12}, the competition between both \cite{szolnoki_pre11b}, and for reward \cite{szolnoki_epl10}. In general, the complexity is largely due to the spontaneous emergence of cycling dominance between the competing strategies, which can manifest in strikingly different ways. By pool-punishment, for example, if the value of $r$ is within an appropriate range \cite{szolnoki_pre11}, the pool-punishers can outperform defectors who in turn outperform cooperators who in turn outperform the pool-punishers, thus closing the loop of dominance. Interestingly, in the absence of defectors peer-punishers and pure cooperators receive the same payoff, and hence their evolution becomes equivalent to that of the voter model \cite{liggett_85}. Notably however, the logarithmically slow coarsening can be effectively accelerated by adding defectors via rare random mutations \cite{helbing_pre10c}. Similarly complex solutions can be observed for rewarding \cite{szolnoki_epl10}. There, if rewards are too high, defectors can survive by means of cyclic dominance, but in special parameter regions rewarding cooperators can prevail over cooperators through an indirect territorial battle with defectors, qualitatively identically as reported for peer-punishment \cite{helbing_ploscb10}. Figure~\ref{indirect} features two sequences of snapshots that demonstrate both evolutionary scenarios. Altogether, these results indicate that second-order free-riding \cite{panchanathan_n04, fehr_n04, fowler_n05b}, referring to cooperators who refrain either from punishing or rewarding, finds a natural solution on structured populations that is due to pattern formation. The aptness of structured populations for explaining the stability and effectiveness of punishment can in fact be upgraded further by means of coevolution \cite{perc_njp12}, as we will review in Sec.~4. On the contrary, while experiments attest to the effectiveness of both punishment \cite{fehr_aer00} and reward \cite{rand_s09} for elevating collaborative efforts, the stability of such actions in well-mixed populations is rather elusive, as reviewed comprehensively in \cite{sigmund_tee07}.

\subsection*{2.4. Statistical physics: avoiding pitfalls}
Before concluding this section and devoting our attention to more complex interaction networks and coevolutionary models, it is important to emphasize difficulties and pitfalls that are frequently associated with simulations of three or more competing, possibly cyclically dominating, strategies on structured populations. Here methods of statistical physics, in particular that of Monte Carlo simulations \cite{liggett_85, binder_88, landau_00}, are invaluable for a correct treatment. Foremost, it is important to choose a sufficiently large system size and to use long enough relaxation times. If these conditions are not met the simulations can yield incorrect one- and/or two-strategy solutions that are unstable against the introduction of a group of mutants. For example, the homogeneous phase of cooperators or pool-punishers can be invaded completely by the offspring of a single defector inserted into the system at sufficiently low values of $r$ \cite{szolnoki_pre11}. At the same time, defectors can be invaded by a single group of pool-punishers (or cooperators) if initially they form a sufficiently large compact cluster. In such cases the competition between two homogeneous phases can be characterised by the average velocity of the invasion fronts separating the two spatial solutions. Note that a system with three (or more) strategies has a large number of possible solutions because all the solutions of each subsystem (comprising only a subset of all the original strategies) are also solutions of the whole system \cite{szabo_pr07}. In such situations the most stable solution can be deduced by performing a systematic check of stability between all the possible pairs of subsystem solutions that are separated by an interface in the spatial system. Fortunately, this analysis can be performed simultaneously if one chooses a suitable patchy structure of subsystem solutions where all relevant interfaces are present. The whole grid is then divided into several domains with different initial strategy distributions containing one, two or three strategies. Moreover, the strategy adoptions across the interfaces are initially forbidden for a sufficiently long initialisation period. By using this approach one can avoid the difficulties associated either with the fast transients from a random initial state or with the different time scales that characterize the formation of possible subsystem solutions. It is easy to see that a random initial state may not necessarily offer equal chances for every solution to emerge. Only if the system size is large enough the solutions can form locally, and the most stable one can subsequently invade the whole system. At small system sizes, however, only those solutions whose characteristic formation times are short enough can evolve. The seminal works considering punishment on structured populations \cite{brandt_prsb03, nakamaru_eer05}, as well as a most recently anti-social punishment \cite{rand_jtb10}, could potentially benefit from such an approach, as it could reveal additional stable solutions beyond the well-mixed approximation \cite{sigmund_pnas01, hauert_s02, hauert_bt08, rand_nc11}.

\begin{figure}
\begin{center} \includegraphics[width = 8.5cm]{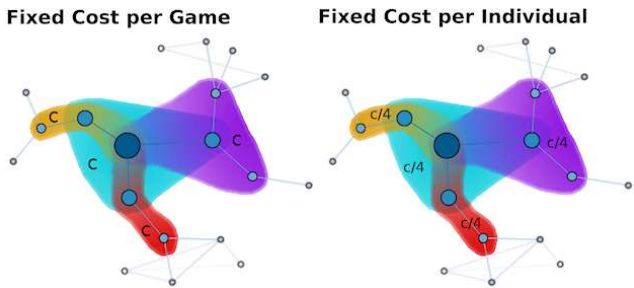}
\caption{When the public goods game is staged on a complex network, cooperators can either bare a fixed cost per game $z$ (left panel), or this cost can be normalized with the number of interactions, i.e., $z/(k_i+1)$, where $k_i$ is the number of neighbors of each particular cooperator $i$. In the latter case, one effectively recovers a fixed cost per individual (right panel). This distinction has significant consequences for the evolution of public cooperation on complex interaction networks, as originally reported in \cite{santos_n08}. Only if the cost is normalized with the number of neighbors does social heterogeneity significantly promote the evolution of public cooperation.}
\label{fig_pacheco}
\end{center}
\end{figure}

\section*{3. COMPLEX NETWORKS}
With the maturity of methods of statistical physics, the availability of vast amounts of digitised data, and the computational capabilities to process them efficiently, it has become possible to determine the actual contact patterns across various socio-technical networks \cite{albert_rmp02, newman_siamr03, boccaletti_pr06}. These studies have shown that the degree distribution $P(k)$ of most real-world networks is highly skewed, and that most of the time it follows a power law $P(k)\sim k^{-\gamma}$ \cite{clauset_siam09}. The heterogeneity of degrees leads to social diversity, which has important consequences for the evolution of cooperation. Although many seminal works concerning evolutionary games on networks have focused on pairwise interactions \cite{szabo_pr07, roca_plr09}, games governed by group interactions are rapidly gaining on popularity.

\subsection*{3.1. Social heterogeneity}

Due to the overwhelming evidence indicating that social heterogeneity promotes the evolution of cooperation in pairwise social dilemma games \cite{santos_prl05, santos_prsb06, santos_pnas06, gomez-gardenes_prl07, gomez-gardenes_jtb08}, it is natural to ask what is its impact on games governed by group interactions. Santos et al.~\cite{santos_n08} have therefore reformulated the public goods game to be staged on complex networks. Every player $i$ plays $k_i+1$ public goods games, as described before for lattices, only that here the degree $k_i$ of every player can be very different. Since the groups will thus also have different size, cooperators can contribute either a fixed amount per game $c_i=z$, or a fixed amount per member of the group $c_i=z/(k_i+1)$, as depicted in Fig.~\ref{fig_pacheco}. Identical to the traditional setup, the contributions within different groups are multiplied by $r$ and accumulated. However, the payoff of an otherwise identical player is not the same for the two different options. By defining the adjacency matrix of the network as $A_{ij}=1$ when individuals $i$ and $j$ are connected and $A_{ij}=0$ otherwise, we obtain the following net benefit $P_i$ for both versions of the game
\begin{widetext}
\begin{equation}
P_{i}=\sum_{j=1}^{N}A_{ij}\frac{r(\sum^{N}_{l=1}A_{jl}s_{l}c_l +s_jc_j)}{k_{j}+1}
+\frac{r(\sum_{j=1}^N A_{ij}s_{j}c_j +s_{i}c_i)}{k_{i}+1}-(k_{i}+1)s_{i}c_i\;,
\label{netpayoff}
\end{equation}
\end{widetext}
where however the precise value of $c_i$ is set depending on whether cooperators bare a fixed cost per game or a fixed cost per player. After each full round of the game all players decide synchronously whether or not they will change their strategy. This is done by following the finite population analogue of the replicator rule. An individual $i$ with payoff $P_i$ randomly selects one neighbor $j$ amongst its $k_i$ contacts. If $P_i \geq P_j$ nothing changes, but if $P_i<P_j$ player $i$ adopts the strategy of the more successful neighbor $j$ with a probability that depends on the difference $\Delta P=P_j-P_i$.

Results presented in \cite{santos_n08} show that heterogeneous networks promote the evolution of public cooperation. Yet this is particularly true when cooperators pay a fixed cost per individual. Cooperation is then viable already at $\eta=r/(\langle k\rangle+1)=0.3$ (normalized multiplication factor), which is less than half of the critical value obtained on lattices. Moreover, heterogeneous networks enable complete cooperator dominance well before cooperative behavior even emerges on regular networks. Phenomenologically, the promotion of cooperation is due to the diversity of investments, which is a direct consequence of the heterogeneity of the underlying network. As cooperators pay a cost that depends on their degree, namely $c/(k+1)$, the fitness landscape becomes very rich and diverse -- a feature absent for lattices. In fact, for a single public goods game the difference between the payoff of a cooperator and defector is no longer proportional to $c$, but rather inversely proportional to the number of games each player plays. This gives and evolutionary advantage to cooperative hubs, i.e., players with a high degree.

The seminal study by Santos et al.~\cite{santos_n08} motivated many others to study the evolution  of public cooperation on complex networks. As evidenced by preceding works considering pairwise social dilemmas, the degree distribution is not the only property that affects the outcome of an evolutionary process \cite{abramson_pre01, ohtsuki_n06, pusch_pre08, assenza_pre08, devlin_pre09b}. Other properties, like the average path length, the clustering coefficient, or the presence of correlations amongst high-degree nodes can be just as important \cite{szabo_pr07}. Rong and Wu \cite{rong_epl09} have explored how the presence of degree correlations affects the evolution of public cooperation on scale-free networks. They found that assortative networks -- those in which alike nodes are likely connected to each other -- act detrimental as heterogeneity no longer confers a natural advantage to cooperative hubs. Conversely, if players with dissimilar degree are more likely connected, the onset of cooperation occurs at lower values of $r$. Similarly, Rong et al.~\cite{rong_pre10b} have investigated the evolution of public cooperation on highly clustered heterogeneous networks, discovering that clustering has a beneficial effect on the evolution of cooperation as it favors the formation and stability of compact cooperative clusters. Yang et al.~\cite{yang_hx_pre09}, on the other hand, adopted a different approach by trying to optimise the number of cooperative individuals on uncorrelated heterogeneous networks. They have considered a variation of the original model \cite{santos_n08}, in which potential strategy donors are no longer chosen randomly but rather proportionally to their degree. It was shown that the promotion of cooperation is optimal if the selection of neighbors is linearly proportional to their degree. While these results indicate that correlations are very important for the evolution of public cooperation, further explorations are needed to fully understand all the details of results presented in \cite{rong_epl09, rong_pre10b, yang_hx_pre09}, which we have here omitted.

We end this subsection by revisiting the role of heterogeneities in the dynamics of investments and payoff distributions, as reviewed before in subsection 2.2. Unlike lattices, complex networks make it interesting to correlate the degree of players with either (i) the investments they make as cooperators \cite{cao_xb_pa10, li_j_pa12}, (ii) the payoffs they are receiving from each group \cite{zhang_hf_pa10, peng_d_epjb10}, or with (iii) both (i) and (ii) together \cite{lei_c_pa10}. These studies exploit the heterogeneity of scale-free networks to implement degree-based policies aimed at promoting cooperation. In \cite{cao_xb_pa10}, for example, it has been shown that positively correlating the contributions of cooperators with their degree acts strongly detrimental on the evolution of public cooperation. On the other hand, if cooperators with only a few connections are those contributing the most, cooperation is promoted. An opposite relation has been established with respect to the correlations between the degree of players and the allocation of payoffs \cite{zhang_hf_pa10, peng_d_epjb10}. In particular, cooperation thrives if players with the highest degree receive the biggest share of the payoff within each group. Moreover, the impact of degree-correlated aspiration levels has also been studied \cite{yang_hx_pa12}, and it was shown that a positive correlation, such that the larger the degree of a player the higher its aspiration level, promotes cooperation. Together, these results indicate that favoring hubs by either decreasing their investments or increasing their payoffs or aspiration promotes the evolution of public cooperation, which in turn strengthens the importance of hubs as declared already in the seminal paper by Santos et al.~\cite{santos_n08}.

\subsection*{3.2. Accounting for group structure: bipartite graphs}

\begin{figure*}
\begin{center}
\includegraphics[width = 14.2cm]{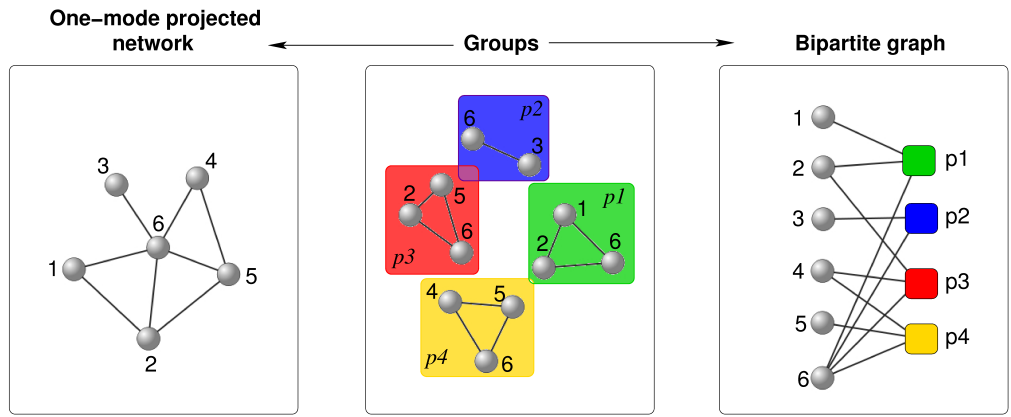}
\caption{Schematic presentation of the two different forms of encoding collaboration data. In the central plot several collaborating groups represent the original data. The interactions amongst players can be translated into a projected complex network (left). However, if one aims at preserving all the information about the group structure, a representation as a bipartite graph (right) is more appropriate. Figure reproduced with permission from \cite{gomez-gardenes_c11}.}
\label{fig_chaos1}
\end{center}
\end{figure*}

The implementation of the public goods game as introduced in \cite{santos_n08} makes an important assumption regarding the composition of groups in which the games take place. This assumption relies on the fact that each group is defined solely on the basis of connections making up the complex interaction network. However, it is rather unrealistic that this definition holds in real social networks, such as for example collaboration networks \cite{newman_pnas04}. Figure~\ref{fig_chaos1} features a schematic presentation of this situation. Suppose we know the actual interaction structure of a system composed of $6$ individuals performing collaborative tasks arranged into $4$ groups (central panel). If we merge this structure into a projected (one-mode) complex network, the collection of groups is transformed into a star-like graph (left panel) having a central hub (node $6$) with five neighbors. By making this coarse-graining, we have lost all the information about the group structure of the system, and it is easy to realise that following \cite{santos_n08} to construct the groups we recover a very different composition made up of $6$ groups of sizes $6$, $4$ ($2$), $3$ ($2$) and $2$, respectively. Moreover, it is important to note that a scale-free distribution of interactions $P(k)\sim k^{-\gamma}$ maps directly to a scale-free distribution of group sizes $P(g)\sim g^{-\gamma}$. However, in reality individuals tend to perform collaborative tasks in groups of a rather homogeneous size \cite{ramasco_pre04}, regardless of the size of the set of their overall collaborators. Accordingly, the distribution of group size is better described by an exponential distribution $P(g)\sim \exp(-\alpha g)$.

To preserve information about both the structure of pairwise ties and the structure of groups, G\'omez-Garde\~nes et al. have introduced the use of bipartite graphs \cite{gomez-gardenes_c11, gomez-gardenes_epl11}. A bipartite representation, as depicted in the right panel of Fig.~\ref{fig_chaos1}, contains two types of nodes. One denoting individuals (circular nodes), and the other denoting groups (square nodes), whereas links connect them as appropriate. Such a bipartite framework is well-suited for studying dynamical processes involving $N$-player interactions.

The setup of the public goods game on bipartite networks is similar to that on one-mode networks, with deviations as described in \cite{gomez-gardenes_c11, gomez-gardenes_epl11}. The graph is composed of $N$ agents playing the game within $G$ (not necessarily equal to $N$) groups whose connections are encoded in a $G\times N$ matrix $B_{ij}$. The $i$-th row of this matrix accounts for all the individuals belonging to group $i$, so that $B_{ij}=1$ when agent $j$ participates in group $i$ while $B_{ij}=0$ otherwise. Alternatively, the information in the $i$-th column encodes all the groups containing agent $i$, i.e., $B_{ji}=1$ when agent $i$ participates in group $j$ and $B_{ji}=0$ otherwise. At each time step player $i$ plays a round of the game in every group it is member. The total payoff after being involved in $q_i=\sum_{j=1}^{G}B_{ji}$ groups can be expressed as
\begin{equation}
P_{i}=\sum_{j=1}^{G}\frac{rB_{ji}}{m_{j}}\left[\sum^{N}_{l=1}B_{jl}s_{l}c_l \right]-s_{i}c_{i}q_{i}\;,
\label{bippayoff}
\end{equation}
where $m_{j}=\sum_{i=1}^{N}B_{ji}$ is the number of individuals in group $j$. Although in principle one could take further advantage of the group structure in order to define different scenarios for the update of strategies, the evolutionary dynamics is defined identically as on one-mode projected networks \cite{gomez-gardenes_c11, gomez-gardenes_epl11}. The updating can rely on the usage of a replicator-like rule \cite{santos_n08}, or the Fermi rule introduced in Eq.~\ref{eq:Fermi}.

\begin{figure}
\begin{center} \includegraphics[width = 7.1cm]{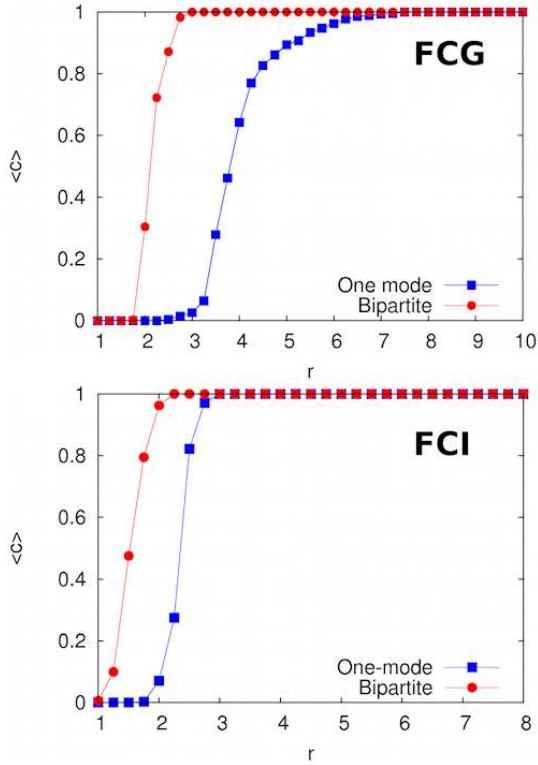}
\caption{Cooperation level $\langle C\rangle$ as a function of the multiplication factor $r$ for the public goods game played on the one-mode (projected) collaboration network and the bipartite graph preserving the original group structure.  The two plots are for the public goods game played in the fixed cost per game (FCG) and the fixed cost per individual (FCI) mode. The strategy updating makes use of the Fermi function (see Eq.~\ref{eq:Fermi}). This figure is reproduced with permission from \cite{gomez-gardenes_c11}.}
\label{fig_chaos2}
\end{center}
\end{figure}

Results presented in \cite{gomez-gardenes_c11} indicate that, regardless of the update rule and the details of the public goods game, the actual group structure of collaboration networks promotes the evolution of cooperation. One arrives to this conclusion by comparing the cooperation level on the bipartite representation of a real collaboration network (containing author-article links) with the  cooperation level on a projected one-mode network that is composed solely of author-author ties (see Fig.~\ref{fig_chaos2}). On the other hand, by comparing the performance of two bipartite structures having different social connectivity -- one having scale-free and the other a Poissonian distribution of degree -- but the same group structure \cite{gomez-gardenes_epl11}, we find that it is the group structure rather than the distribution of degree that determines the evolution of public cooperation. In particular, the promotion of cooperation due to a scale-free distribution of degree as reported in \cite{santos_n08} is hindered when the group structure is disentangled from the social network of contacts by means of the bipartite formulation.

Notably, the bipartite formulation has recently been revisited by Pe\~na and Rochat \cite{pena_pone12}, who compared the impact of different distributions used separately for group sizes and the number of individual contacts. They showed that a key factor that drives cooperation on bipartite networks is the degree of overlap between the groups. The later can be interpreted as the bipartite analogue of the clustering coefficient in one-mode networks, which as reviewed above, is highly beneficial for the evolution of public cooperation. The results reported in \cite{pena_pone12} also help to understand the high level of cooperation observed in one-mode scale-free networks \cite{santos_n08}, because the assumption that the structure of groups is implicitly defined by the network itself imposes a high degree of overlap between the groups, especially for scale-free networks.

\subsection*{3.3. Other network-based frameworks}
\begin{figure}
\begin{center}
\includegraphics[width = 4.0cm]{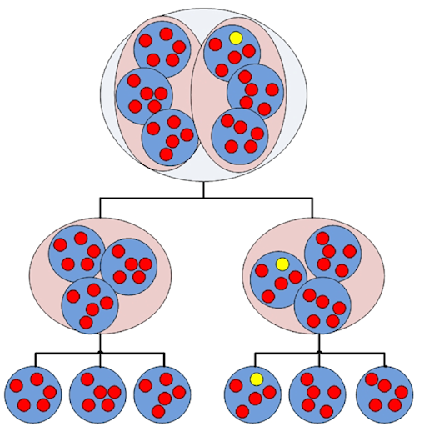}
\\ \vspace{0.6cm}
\includegraphics[width = 5.0cm]{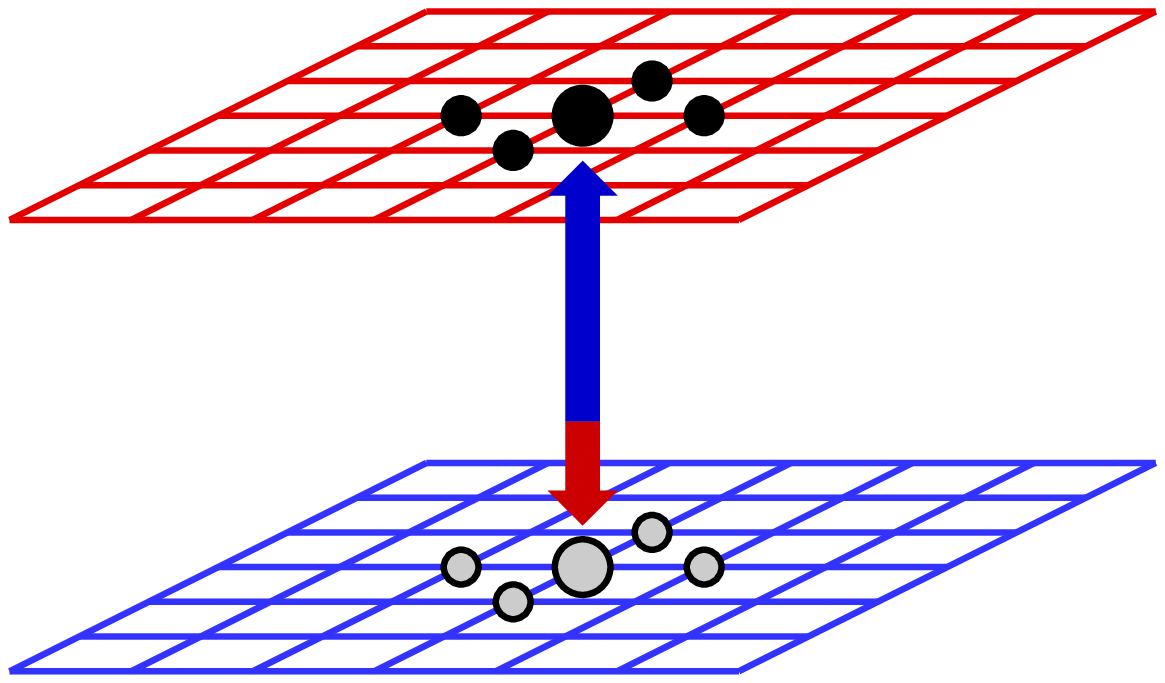}
\caption{The upper panel shows the multilevel hierarchical structure introduced in \cite{wang_j_epl11}, where groups are hierarchically ordered as modules of a network in which the public goods game is played. This figure is reproduced with permission from \cite{wang_j_epl11}. The bottom panel shows two interdependent lattices as studied in \cite{wang_z_epl12}. Players can adopt different strategies within each layer, but coupling between the payoffs obtained in each of the two layers (see Eq.~\ref{coupling}) makes their evolutionary dynamics interdependent.}
\label{fig_other}
\end{center}
\end{figure}

In addition to the distributions of individual contacts and group sizes, the impact of other topological features of social networks has also been studied. In particular, in \cite{wang_j_epl11} the authors studied a hierarchical social structure composed of communities or modules in which several public goods games are played simultaneously. For a setup with $2$ hierarchical levels we thus have the following framework: player $i$ is member in one group of size $m$ at the lowest level and, simultaneously, it is also a member in a larger group together with the rest of the population. This setup can be generalised to systems composed of $n$ hierarchical levels, as shown in the upper panel of Fig.~\ref{fig_other} for $n=3$. The coupling between the evolutionary dynamics in each of the levels is accomplished by splitting the contribution $c$ of each cooperator by the number of groups, and by choosing a different probability for the updating rules within and between modules. Results reported in \cite{wang_j_epl11} indicate that public cooperation is promoted when imitation between players belonging to different modules is strong, while at the same time the imitation between players within the same lowest level module is weak. This combination of strengths leads to the onset of groups composed solely of cooperators, but it also enables cooperators that coexist with defectors to avoid extinction.

Another important structural feature recently addressed is multiplexity \cite{buldyrev_n10, gao_jx_np12, parshani_prl10}, or the coupling between several network substrates. Although this structural ingredient has only recently been tackled in the field of network science, some works on the subject have already appeared in the context of evolutionary games \cite{gomez-gardenes_srep12, wang_z_epl12}. In \cite{wang_z_epl12}, where the focus was on group interactions, the authors have studied a simple layered framework in which two regular lattices were coupled as depicted schematically in the bottom panel of Fig.~\ref{fig_other}. The rationale is that a given individual is represented in each of the two layers, although in principle, it can adopt different strategies in each of them. The coupling between layers is solely due to the utility function, which couples the payoff $P_i^A$ obtained on layer $A$ and the payoff $P_j^B$ obtained on layer $B$ as
\begin{equation}
U_i=\alpha P_i^A+(1-\alpha)P_i^B,\,\,
U_j=(1-\alpha) P_j^B+\alpha P_i^A\,.
\label{coupling}
\end{equation}
The parameter $\alpha\in[0,1]$ determines the bias in each layer. When $\alpha \rightarrow 0$ ($\alpha\rightarrow 1$) the dynamics of layer $A$ ($B$) is almost fully driven by layer $B$ ($A$). An intriguing result reported in \cite{wang_z_epl12} is that as one layer almost dominates the other, cooperation is very much favored in the slave layer, i.e., in $A$ when $\alpha \rightarrow 0$ or in $B$ when $\alpha \rightarrow 1$. Obviously, the master layer then behaves almost equally as an isolated graph, showing a greater vulnerability to defection than the slave layer. These initial results invite further research concerning the impact of multiplexity of social networks on the evolution of cooperation.

\subsection*{3.4. Populations of mobile agents}
Prior to focusing on coevolutionary rules, we review one special case in which a population of mobile players is embedded in a physical space so that a time-varying network of interactions is constructed sequentially and in accordance with their movements. Two possible scenarios must be distinguished. First, there are studies in which the movement of players is independent of the evolutionary dynamics \cite{meloni_pre09, chen_z_pa11b, zhang_j_pa11, cardillo_pre12}. Effectively, the movements thus correspond to a random walk. Second, the motion of players can be affected by the outcome of the game \cite{helbing_pnas09, jiang_ll_pre10, yang_hx_pre10, lin_yt_pa11, roca_pnas11, cong_pone12, xia_cy_acs12}. In addition to this classification, we must also distinguish two different types of space in which the players live. In particular, players can either move on a lattice, or they can move across continuous space. In the former case, the network of interactions is set simply by considering two players occupying two adjacent sites as connected, so that the resulting graph is a square lattice with a certain fraction of missing links \cite{chen_xj_pre12b}. The usage of continuous space, on the other hand, requires the construction of a random geometric graph every time after all the players have changed their position. Such graphs are typically constructed by connecting together all pairs of players that are less apart than a given threshold $R$. This introduces an additional parameter that allows an interpolation between a fully unconnected ($R=0$) and a fully connected ($R \rightarrow \infty$) graph.

An interesting study where the movements were uncorrelated with the evolutionary dynamics was performed by \cite{cardillo_pre12}. The players moved randomly with a constant velocity $v$ in a continuous 2-dimensional plane, establishing a random geometric graph with constant radius $R$ after all agents have made a move. The groups in which the public goods game was played where then constructed as introduced in \cite{santos_n08}. Two resonance-like phenomena were reported, given that the fraction of cooperators exhibited a bell-shaped dependence on both $v$ and $R$. Accordingly, an intermediate degree of mobility, as well as an intermediate level of connectedness amongst the mobile players were found to be optimal for the evolution of cooperation. The maximum was found to be closely related to the percolation threshold of a random geometric graph, much in agreement with preceding results on static networks \cite{wang_z_pre12b}.

The setup where the mobility was driven by the evolutionary dynamics was explored more often, especially for games governed by pairwise interaction. Factors that can affect how and when the players move include their fitness \cite{helbing_pnas09, jiang_ll_pre10}, aspiration level \cite{yang_hx_pre10, lin_yt_pa11}, as well as reputation \cite{cong_pone12}. In terms of group interactions, Roca et al.~\cite{roca_pnas11} considered a system of $N$ agents occupying a $L \times L > N$  square lattice. The players were allowed to move to an empty site if their aspirations were not met. They have showed that only moderate greediness leads to high levels of public cooperation and social agglomeration. A similar model was studied by Xia et al.~\cite{xia_cy_acs12}, who showed that the provisioning of local information about the payoffs of nearest neighbors does not alter the original conclusions presented in \cite{roca_pnas11}.

We conclude this subsection by noting that payoff-driven mobility has also been explored in the framework of metapopulations \cite{zhang_jl_csf11}. There a population of $N$ players moves across a network of $M$ nodes, where $M>N$. Thus, when several players meet on the same node of the network, they play a round of the public goods game. Subsequently, based on the difference between the collected payoff and their aspiration level, they decide whether to stay or to move to a neighboring node. An interesting result reported in \cite{zhang_jl_csf11} is that the larger the ratio $M/N$, and hence the larger the average size of groups in which the game is played, the better the chances of cooperators to survive.

\section*{4. COEVOLUTIONARY RULES}
Coevolutionary models go beyond structured populations in the sense that the interaction network itself may be subject to evolution \cite{wu_t_epl09, wu_t_pre09, zhang_hf_epl11, zhang_cy_epjb11}. However, this need not always be the case, as the coevolutionary process can also affect system properties other than the interaction network, like for example the group size \cite{janssen_jtb06}, heritability \cite{liu_rr_pa10}, the selection of opponents \cite{shi_dm_pa09}, the allocation of investments \cite{zhang_hf_pa12}, the distribution of public goods \cite{perc_pre11}, or the punishment activity of individual players \cite{perc_njp12}. Possibilities seem endless, as recently reviewed for games governed by pairwise interactions \cite{perc_bs10}. Games governed by group interactions have received comparatively little attention.

\begin{figure}
\begin{center}
\includegraphics[width = 8.5cm]{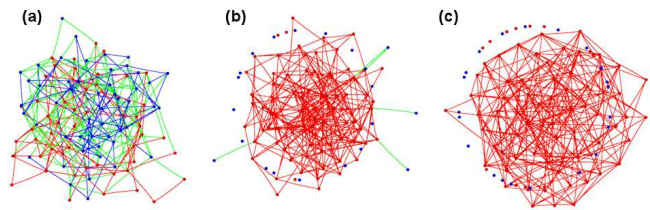}
\caption{Coevolution of strategy and structure leads to high levels of public cooperation. Networks depict snapshots in time at $0$ (a), $2000$ (b) and $20000$ (c) iterations, whereby green links connect defector-cooperator pairs, blue links connect two defectors, while red links connect two cooperators. Accordingly, cooperators are depicted red and defectors are depicted blue. This figure was reproduced with permission from \cite{wu_t_epl09}, where also further details with respect to the simulation setup can be found.}
\label{nasty}
\end{center}
\end{figure}

One of the earliest coevolutionary rules affecting the interaction network during a public goods game was proposed and studied by Wu et al. \cite{wu_t_epl09}, where it was shown that adjusting the social ties based on the payoffs of players may significantly promote cooperation. If given an opportunity to avoid predominantly defective groups (referred to as a ``nasty environment''), the population can arrive at a globally cooperative state even for low values of $r$. Interestingly, decoupling the coevolutionary adjustment of social ties with the evolution of strategies renders the proposed rule ineffective in terms of promoting public cooperation. As depicted in Fig.~\ref{nasty}, allowing for the coevolution of strategy and structure leads to predominantly cooperative states out of an initially mixed population of cooperators and defectors.

Alternative coevolutionary models affecting the interactions amongst players were also studied in \cite{wu_t_pre09, zhang_hf_epl11, zhang_cy_epjb11}, with the prevailing conclusion being that the evolution of public cooperation can benefit greatly from the interplay between strategy and structure. In particular, aspiration-induced reconnection can induce a negative feedback effect that stops the downfall of cooperators at low values of $r$ and lead to heterogeneous interaction networks \cite{zhang_hf_epl11}, while strategy-inspired additions and deletions of links between players can lead to hierarchical clustering \cite{zhang_cy_epjb11}. Also worth noting is the first coevolutionary model making use of the bipartite network formalism \cite{smaldino_pone11}, where individuals can switch groups. An implementation of social policies is thus possible and in \cite{smaldino_pone11} it was shown that restricting the maximum capacity of groups is a good policy for promoting cooperation.

Another interesting coevolutionary rule is the preferential selection of opponents introduced in \cite{shi_dm_pa09}. It was shown that a simple payoff-based selection can lead to higher payoffs along the boundaries separating cooperators and defectors, which in turn facilitates spatial reciprocity and leads to larger cooperative clusters. Likewise leaving the interactions amongst players unchanged is the dynamic allocation of investments \cite{zhang_hf_pa12} and the success-driven distribution of public goods studied in \cite{perc_pre11}. Both frameworks have the ability to promote cooperation, although in the latter case the complete dominance of cooperators may be elusive due to the spontaneous emergence of super-persistent defectors. This in turn indicates that success-driven mechanisms are crucial for effectively harvesting benefits from collective actions, but that they may also account for the observed persistence of maladaptive behavior.

\begin{figure}
\begin{center} \includegraphics[width = 8.5cm]{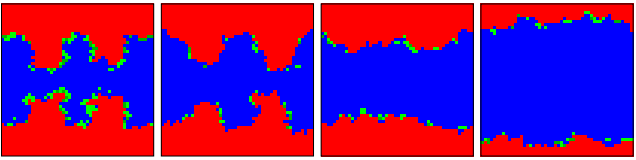}
\caption{\label{adaptive} Rough interfaces enable defectors (red) an effective exploitation of cooperators (blue), thus hindering spatial reciprocity. Upon the introduction of adaptive punishment (green, where darker (lighter) shades imply stronger (weaker) punishing activity), interfaces become smoother, which in turn invigorates spatial reciprocity and prevents defectors from being able to exploit the public goods. A prepared initial state, corresponding to a rough interface, is used to reveal the workings of this mechanism. Interestingly, here the stationary state is a pure $C$ phase, while under the same conditions peer-punishment without coevolution yields a pure $D$ phase. We refer to \cite{perc_njp12} for further details.}
\end{center}
\end{figure}

Strategic complexity may also be subject to coevolution, as proposed and studied in \cite{perc_njp12}, where players were allowed to adapt their sanctioning efforts depending on the failure of cooperation in groups where they were members. Preceding models assumed that, once set, the fine and cost of punishment do not change over time \cite{helbing_ploscb10}. However, by relaxing this restriction one obtains the spontaneous emergence of punishment so that both defectors and those unwilling to punish them with globally negligible investments are deterred. Crucial is the fact that adaptive punishers are able to smooth the interfaces between cooperators and defectors, as demonstrated in Fig.~\ref{adaptive}. This indicates that coevolution may be the key to understanding complex social behavior as well as its stability in the presence of seemingly more cost-efficient strategies.

Before concluding this review, it is important to point out that diffusion may also be seen as a coevolutionary process \cite{perc_bs10}, as it allows players to move in the population. In a series of papers, Wakano \textit{et al.} \cite{wakano_jtb07, wakano_pnas09, wakano_jtb11} have elaborated extensively on the patterns that may arise in two-dimensional continuous space. A detailed analysis of the spatiotemporal patterns based on the Fourier analysis and Lyapunov exponents reveals the presence of spatiotemporal chaos \cite{wakano_jtb11}, which fits to the complexity of solutions one is likely to encounter when studying group interactions on structured populations.

\section*{5. OUTLOOK}
Although our understanding of evolutionary processes that are governed by group interactions has reached a remarkably high level, there still exist unexplored problems that require further attention. While physics-inspired studies account for the majority of recent advances on this topic \cite{szabo_pr07, roca_plr09, perc_bs10}, there also exist many experimental and theoretical results on well-mixed populations that would be interesting to verify on structured populations.

The ``stick versus carrot'' dilemma \cite{andreoni_aer03, hilbe_prsb10, gneezy_prsb12}, for example, is yet to be settled on structured populations. It is also important to note that recent research related to antisocial punishment \cite{herrmann_s08, nakamaru_jtb06, rand_jtb10, rand_nc11} and reward \cite{sigmund_pnas01, andreoni_aer03, rand_s09, hilbe_prsb10, szolnoki_epl10} is questioning the aptness of sanctioning for elevating collaborative efforts and raising social welfare. The majority of previous studies addressing the ``stick versus carrot'' dilemma concluded that punishment is more effective than reward in sustaining public cooperation \cite{sigmund_tee07, hauert_jtb10}. However, evidences point out that rewards may be as effective as punishment and lead to higher total earnings without potential damage to reputation \cite{milinski_n02, semmann_bes04} or fear from retaliation \cite{dreber_n08}. In view of recent advances concerning punishment \cite{helbing_ploscb10, helbing_pre10c, helbing_njp10, szolnoki_pre11, szolnoki_pre11b} and reward \cite{szolnoki_epl10} on lattices, it seems worth continuing in this direction also with antisocial punishment and the competition between punishment and reward in general. There is also the question of scale at which social dilemmas are best resolved \cite{santos_pnas11}, as well as the issue of the emergence of fairness in group interactions \cite{van-segbroeck_prl12}, which could also both be examined on structured populations.

Complex interactions networks also offer many possibilities for future research on games governed by group interactions. The concept of bipartitness \cite{gomez-gardenes_c11, gomez-gardenes_epl11}, for example, appears to be related to multilevel selection \cite{traulsen_pnas06, wang_j_epl11}, which however was so far considered without explicit network structure describing the interactions amongst players. Motivation can also be gathered from coevolutionary games \cite{perc_bs10}, where group interactions on structured populations can still be considered as being at an early stage of development. While initially many studies that were performed only for pairwise social dilemmas appeared to be trivially valid also for games that are governed by group interactions, recent research has made it clear that at least by default this is in fact not the case. In this sense the incentives are clearly there to reexamine the key findings that were previously reported only for pairwise games on complex and coevolutionary networks also for games that are governed by group interactions.

\section*{6. SUMMARY}
Group interactions on structured populations can be much more than the sum of the corresponding pairwise interactions. Strategic complexity, different public benefit functions, and coevolutionary processes on either lattices or complex networks provide a rich playground that can be explored successfully with methods of statistical physics \cite{albert_rmp02, szabo_pr07, castellano_rmp09}. Research performed thus far offers a thorough understanding of the many key phenomena that can be uniquely associated with games governed by group interactions. On lattices, group interactions effectively link players that are members of the same groups without there being a physical connection between them. This renders local particularities of interaction networks largely unimportant for the outcome of the evolutionary process, and it introduces the deterministic limit of strategy imitation as optimal for the evolution of public cooperation \cite{szolnoki_pre09c}. On the other hand, the size of the group \cite{szolnoki_pre11c} as well as public benefit functions \cite{szolnoki_pre10} gain markedly on significance, thus offering new possibilities for exploration. Strategic complexity \cite{szabo_prl02, helbing_njp10, szolnoki_epl10, szolnoki_pre11, szolnoki_pre11b} significantly increases the complexity of solutions due to spatial pattern formation, yet the obtained results provide elegant explanations for several long-standing problems in the social sciences. Examples include the second-order free-rider problem \cite{helbing_ploscb10}, as well as the stability of reward \cite{szolnoki_epl10} and the successful evolution of institutions \cite{szolnoki_pre11, szolnoki_pre11b}, all of which require additional strategic complexity on well-mixed populations in order to be explained. Complex networks and coevolutionary models further extend the subject with insightful results concerning bipartitness \cite{gomez-gardenes_c11, gomez-gardenes_epl11} and the rewiring of social ties \cite{wu_t_epl09, zhang_hf_epl11}, all adding significantly to our understanding of the provisioning of public goods in human societies.

Although the origins of prosocial behavior in groups of unrelated individuals are difficult to trace down -- there exist evidence indicating that between-group conflicts may have been instrumental for enhancing in-group solidarity \cite{bowles_11}, yet alloparental care and provisioning for someone else's young have also been proposed as viable for igniting the evolution of our other-regarding abilities \cite{hrdy_11} -- it is a fact that cooperation in groups is crucial for the remarkable evolutionary success of the human species, and it is therefore of importance to identify mechanisms that might have spurred its later development \cite{sigmund_10, nowak_11}. The aim of this review is to highlight the importance of such group interactions, and to demonstrate the suitability of methods of statistical physics and network science for studying the evolution of cooperation in games that are governed by them.

\begin{acknowledgments}
This research was supported by the Slovenian Research Agency (Grant J1-4055), the Spanish DGICYT under projects FIS2011-25167 and MTM2009-13848, the Comunidad de Arag\'on through a project to FENOL, and the Hungarian National Research Fund (Grant K-101490). J.G.G. is supported by MEC through the Ram\'on y Cajal program.
\end{acknowledgments}

\end{document}